\documentstyle[epsfig]{article}  

\newcommand{\be}{\begin{equation}}
\newcommand{\ee}{\end{equation}}

\begin{document}

\title
{New phases of QCD;\\ the tricritical point;\\ and RHIC as a ``nutcracker''}  

\author{E.V.Shuryak \\
Department of Physics and Astronomy, State University of New York, \\
     Stony Brook, NY 11794-3800, USA}

\maketitle

\begin{abstract}
It is a talk at 15-th Winter Workshop on Nuclear Dynamics, Parkcity
Utah, Jan.1999.
Because too many interesting things are going on now, I have tried to squeeze
three different subjects into one talk. The first is a brief summary
of the color super-conductivity. During the last year we learned that
instanton-induced forces can not only break chiral symmetry in the
QCD vacuum, but also create correlated scalar diquarks and form new phases,
some similar to the Higgs phase of the Standard model. The second
issue I discuss is the remnant of the so called $tricritical$ point,
which in QCD with physical masses is the endpoint of the first order 
transition. I will argue that exchange of sigmas (which are massless
at this
point even with quark masses included) create interesting
event-by-event
fluctuations, which can be used to locate it. Finally I describe first
results
in $flow$ calculations for non-central collisions at RHIC. It was
found that it is extremely sensitive to Equation of State
(EOS). Furthermore, the unusual ``nutcracker'' picture emerges for 
lattice-motivated EOS, which is formation of two $shells$ which are
physically separated $before$ the freeze-out.

\end{abstract}

\section{QCD at high density}
   New phases of QCD at high density and the color super-conductivity
   issue are a part of  a broader context, the studies of how the
   confining and
chirally asymmetric  QCD ground state is substituted by other phases
as the temperature, the chemical potential,  the number of flavors
(or any combination of those)
are increased. The key player in most of those effects
(except confinement) happen to be instantons, see recent review \cite{SS_98}.   
In the QCD vacuum, for example, the quark condensate is simply the
 density of (almost) zero modes, originating from 
a superposition of zero modes associated with isolated instantons
and anti-instantons.

  At high temperature we expect to find the quark-gluon plasma (QGP) phase
 in which chiral symmetry is restored. So  the 
density of (almost) zero modes  goes to zero.
 This can only be realized if the instanton ensemble changes 
from a nearly random one to a correlated
system with finite clusters, e.g.
  instanton-anti-instanton ($\bar I I$) molecules.
The same is expected to happen (even at T=0) for sufficiently large
number of flavors: in this case the expected next phase is the so
called $conformal$ phase.

 The QCD at finite baryon density 
lay dormant since 70's, when basic applications of QCD like Debye
screening were made.
 It  was revived recently
when it was realized that not only we expect the 
high density phase of QCD to be a color superconductor, as proposed in
\cite{Frau_78,Bar_77,BL_84} with gaps in the MeV range, 
but that the instanton-induced effects lead to much larger
gaps on the order of 100 MeV \cite{RSSV_98,ARW_98}.

 In the next year it was realized that 
the phase structure of QCD at finite baryon density is actually very rich. In 
addition to the dominant order parameter, which is a scalar-isoscalar
color anti-triplet ud diquark, many other condensates form.
 The overall picture
can be characterized by some kind of 
``triality'', both of three major phases under consideration, as well as
of  three competing attractive channels.
These  basic phases are: (i)
the {\em hadronic} (H) phase,
 with (strongly) broken chiral  symmetry
 (ii) the 
{\em color superconductor} (CSC) phase, 
with  broken color symmetry ; and (iii) the {\em quark-gluon   plasma} (QGP) phase, in which
there are no condensates but the instanton ensemble is non-random.
The three basic channels are the instanton-mediated
attraction in (i) $\bar q q$ and (ii) qq channels (responsible for H and CSC phases)
and  the (quark-mediated) attraction between $\bar I I$ ,
 confining the topological charge in the QGP phase.
The interrelation between these three attractive channels and phases
is
not straightforward:
 e.g. the  $<\bar q q>$ may or may not be present in the CSC phase,
and $\bar I I$ molecules have non-zero presence  everywhere. However,
 this paper is still basically about
a competition between these three attractive forces in  different
conditions.     

\begin{figure}[ht]
\epsfig{file=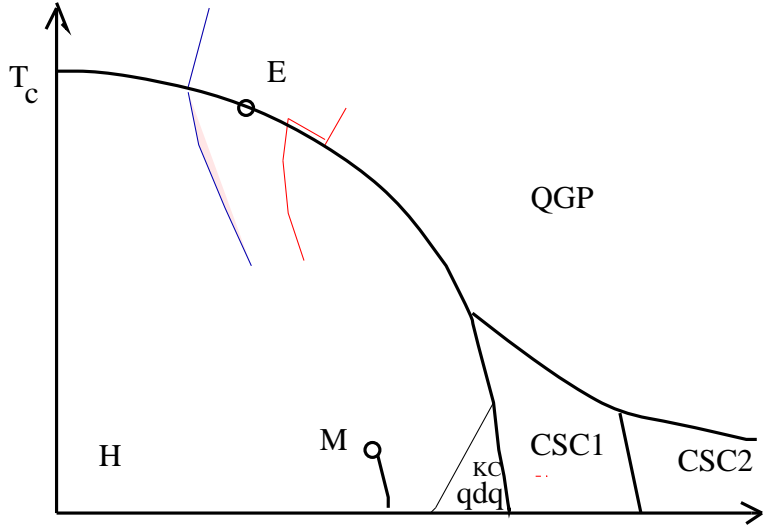,width=7.cm,angle=-90}
\vskip -0.05in
\caption[]{
 \label{shuryak_fig1}
Schematic phase diagram of QCD phases as a function of temperature T
and baryonic chemical potential $\mu$,
as we understand it today. The phases denoted by H and QGP are
the usual hadronic phases (with nonzero $<\bar q q>$ and the
quark-gluon plasma (no condensates). The color super-conducting phases
 CSC1 and
CSC2 have various  $<q q>$ condensates,  the latter has broken
chiral
symmetry and tends asymptotically to color-flavor locking scenario.
KC and QDQ are two  possible phases, with Kaon condensate or
quark-diquark gas.
 E is the endpoint of the 1-st
order transition, M (from multi-fragmentation)  is the endpoint of another 1-st order transition,
between liquid and gas phases of nuclear matter. Two schematic trajectories
corresponding to adiabatic expansion in heavy ion collisions are also
indicated. 
}
\end{figure}

The overview of the situation on the phase diagram is given by
Fig.\ref{shuryak_fig1}, where one can see an approximate location of
color super-conducting phases, as well as
 few schematic trajectories of excited matter, as it expands and
cools
in heavy ion collisions. One may see from those that unfortunately
this new phase region corresponds to rather cool matter,
and so it is $not$ crossed by them. 
Therefore, color super-conductivity
should  only exists in compact stars. This created a challenge,
known as the ``pulsar cooling problem'': a naked Fermi sphere is not
allowed,
because it generates too rapid cooling rate in contradiction to data.

The scenario depends crucially on the number of quark flavors $N_f$.
We start with discussion of (i) $N_f=2$ massless quarks, u,d; (ii)
then move on to  $N_f=3$ massless quarks, and finally to (iii) real
QCD with non-zero quark masses.

{\bf In the $N_f=2$ case} the instanton-induced interaction is 4-fermion
one. Its role in breaking chiral symmetry and making pion light and
$\eta'$ heavy is well documented, see e.g. \cite{SS_98}. One can
Fierz
transform it to diquark channels, which 
contain both color antisymmetric ${ \bar 3}$
and  symmetric $6$ terms. The scalar and the tensor are attractive:
\newpage
$$
\label{l_diq}
{\cal L}_{diq} = 
{g\over 8 N_c^2} \left\{
 -{1\over N_c-1}
 \left[ (\psi^T{\cal F}^T C \tau_2 \lambda_A^a {\cal F}\psi)
        (\bar\psi{\cal F}^\dagger\tau_2 \lambda_A^a C {\cal F}^*\bar\psi^T)  
\right. \right.\nonumber
$$ \be
   \left.\left. \mbox{}
       +(\psi^T{\cal F}^T C \tau_2 \lambda_A^a \gamma_5 {\cal F}\psi)
         (\bar\psi{\cal F}^\dagger \tau_2 \lambda_A^a \gamma_5 C {\cal F}^*
\bar\psi^T) \right]\right. \nonumber\\
\ee $$
  \left. \mbox{}
        +{1\over 2(N_c+1)}
        (\psi^T{\cal F}^T C \tau_2 \lambda_S^a \sigma_{\mu \nu} {\cal F}\psi)
        (\bar\psi{\cal F}^\dagger \tau_2 \lambda_S^a \sigma_{\mu \nu} C 
{\cal F}^*\bar\psi^T) 
        \right\}
$$
where $\tau_2$ is the anti-symmetric 
Pauli matrix, $\lambda_{A,S}$ are the anti-symmetric (color ${ \bar 3}$)
 and symmetric (color {\bf 6}) color generators (normalized in an 
unconventional way,
${\rm}tr(\lambda^a\lambda^b)=N_c\delta^{ab}$, in order to facilitate the 
comparison between mesons and diquarks). As discussed in ref.\cite{RSSV_98}, 
in the  case of two colors 
there is the so called Pauli-G{\"u}rsey symmetry 
 which mixes quarks with anti-quarks. So 
diquarks (baryons of this theory)  are degenerate 
with the corresponding mesons. 
It  manifests itself in the Lagrangians given above: in
this case the coupling constants in $\bar q q$ and qq channels are
the same and
 the
scalar diquarks, like pions. have the mass vanishing in the 
chiral limit. 

Standard BCS-type mean field treatment leads to gap equation, from
which one extract all properties of the color superconductor. Let me
omit details and only mention the bottom line. The $chiral$ symmetry
is
restored, while $color$ SU(3) is broken to SU(2) by the colored
condensate.

{\bf In the $N_f=3$ case}   the situation becomes more interesting.
( Since the critical chemical potential $\mu_c\sim 300-350 MeV$ 
is larger than the strange quark mass $m_s\simeq 140$ MeV, strange 
quarks definitely have to be included.)
 There are several qualitatively new 
features. First, since $N_f=N_c$,
there are new order parameters in which the color and flavor 
orientation of the condensate is locked \cite{ARW_98b}. Second, 
the instanton induced interaction is a four-fermion vertex, so it
does not directly lead to the BCS instability, unless there is also
a $<\bar q q>$ condensate as well. So we need a superconductor where
 chiral symmetry is still $broken$. 
This is indeed what we have found \cite{RSSV2}, after a rather
tedious
calculation. 

{\bf In the $N_f=3$ case with variable strange quark mass}  the
algebraic difficulties increase further. There are dozens of qq and 
$\bar q q$ condensates present, all competing for the resources. 
The largest, ud one, is still in the 100 MeV range, but the smallest
are just few MeV, or comparable with light quark masses. 
Still, those small condensates are enough to solve the ``pulsar
cooling
problem'' (while without strangeness it remained unsolved).

We have found
that two cases discussed above are in fact separated by a first order
transition line, as a function of density or $m_s$. Partially this is
caused
by simple kinematic-al mismatch between $p_F(u,d)$ and  $p_F(s)$
preventing their pairing, if  $m_s$ is large enough.

Finally let me mention that a transition region between nuclear matter
and CSC (see Fig.1)
was claimed before by such exotic phase as Kaon condensation.
In \cite{RSSV2} we propose another (also exotic) quark-diquark (QDQ in Fig.1)
phase, in which
nucleons
dissociate into Fermi gas of constituent quarks plus Bose gas of
constituent 
ud diquarks. 

\section{ Event-by-event fluctuations and possible signatures of the tricritical point  }

  We now discuss the part of the
  phase diagram shown in  Fig.\ref{shuryak_fig1}
for densities below those for color super-conductivity.
At high T and zero density it is believed to be second order if quark masses
and strangeness is ignored, and a simple crossover otherwise. The
discussion of the previous section (and many models 
e.g. the random matrix one
 \cite{HJSV}) suggest  
 that it
 is likely to turn first order at some critical density. 
This means that there should be
 a tri-critical point in the phase diagram with $N_f=2$ massless quarks,
 or the Ising-type endpoint E
if quarks are not massless.
The proposal to search for it experimentally was recently made by
Stephanov, Rajagopal and myself \cite{SRS}. A detailed paper about 
event-by-event fluctuations around this point  \cite{SRS2} is the
basis of this section.

  The main idea is of course based on the existence of truly massless
mode at this point, the sigma field, which is responsible for
``critical opalescence'' and large fluctuations.
The search itself should be partially similar to ``multi-fragmentation''
phenomenon in low energy heavy ion collisions, which is also due to
the endpoint M (see Fig.1)
 of another first order transition. The ``smoking gun'' is
supposed to be a non-monotonous behavior of observable as a function
of such control parameters as collision energy and centrality.
 One can use pions as a ``thermometer'' to measure this fluctuations.

(Note that  in many ways it is the opposite of the 
DCC idea: in that case the pion was the light fluctuating field, while  
its coupling to heavy and wide  and strongly  damping  sigma field is the
main obstacle.)  

We have studied three ways in which sigmas can show up. First and the
simplest
is the ``thermal contact'' idea: at the critical point the sigmas
specific heat becomes large, and this shows up in the pion
fluctuations
just due to energy conservation. The second is ``dynamical exchange'':
pions can exchange the sigmas and this leads to long-range effects,
over
the whole correlation range. Both effects are in 10-20 percent range,
after realistic account for correlation length is made. It is not
large, but much larger than the accuracy of the measurements. 
The third effect is due to sigma decays into  pions, which affect
spectra at small $p_t$ and (even more so) the multiplicity fluctuations.

Large acceptance detectors can 
study the event-by-event (ebe) fluctuations quite easily.
The first data  by NA49 detector at CERN on
 distributions
of $N$, the charged pion multiplicity, and $p_T$
(the mean transverse momentum of the charged pions in an event) 
for
central
 PbPb collisions at 160 AGeV 
display
beautiful Gaussians. 
Since any system in thermodynamic equilibrium exhibits
Gaussian fluctuations, it is natural to ask how much
of the observed fluctuations are thermodynamic in origin \cite{Stodolsky,Shu_fluct}.
We have answered this question quantitatively in this paper,
considering
fluctuations in pion number, mean $p_t$ and their correlation. 
We model the matter at freeze-out as an ideal gas of pions
and resonances in thermal equilibrium, and make quantitative
estimates of the thermodynamic fluctuations in the resulting
pions, many of which come from the decay of the resonances
after freeze-out. The conclusion is that nearly all answers are
reproduced
by the resonance gas, with remaining part likely to be due to
experimental corrections, due to two-particle track resolution and
non-pion admixture. The good agreement
between the non-critical thermodynamic fluctuations
we analyze in Section 3 and NA49 data make it unlikely
that central PbPb collisions at 160 AGeV freeze out
near the critical point.

Estimates  suggest that the critical point is
located at a $\mu_f$ such that it will be found at
an energy between 160 AGeV and AGS energies. This makes it
a prime target for detailed study at the CERN SPS
by comparing data taken at 40 AGeV, 160 AGeV, and in between.
We are more confident
in our ability to describe the properties of the
critical point and thus {\it how} to find it than
we are in our ability to predict where it is.  
If it is located at such a low $\mu$ that the maximum
SPS energy is insufficient to reach it, 
it would
then be in a regime accessible to study by the 
RHIC experiments.  

\begin{figure}[ht]
\epsfxsize=4.in
\epsfig{file=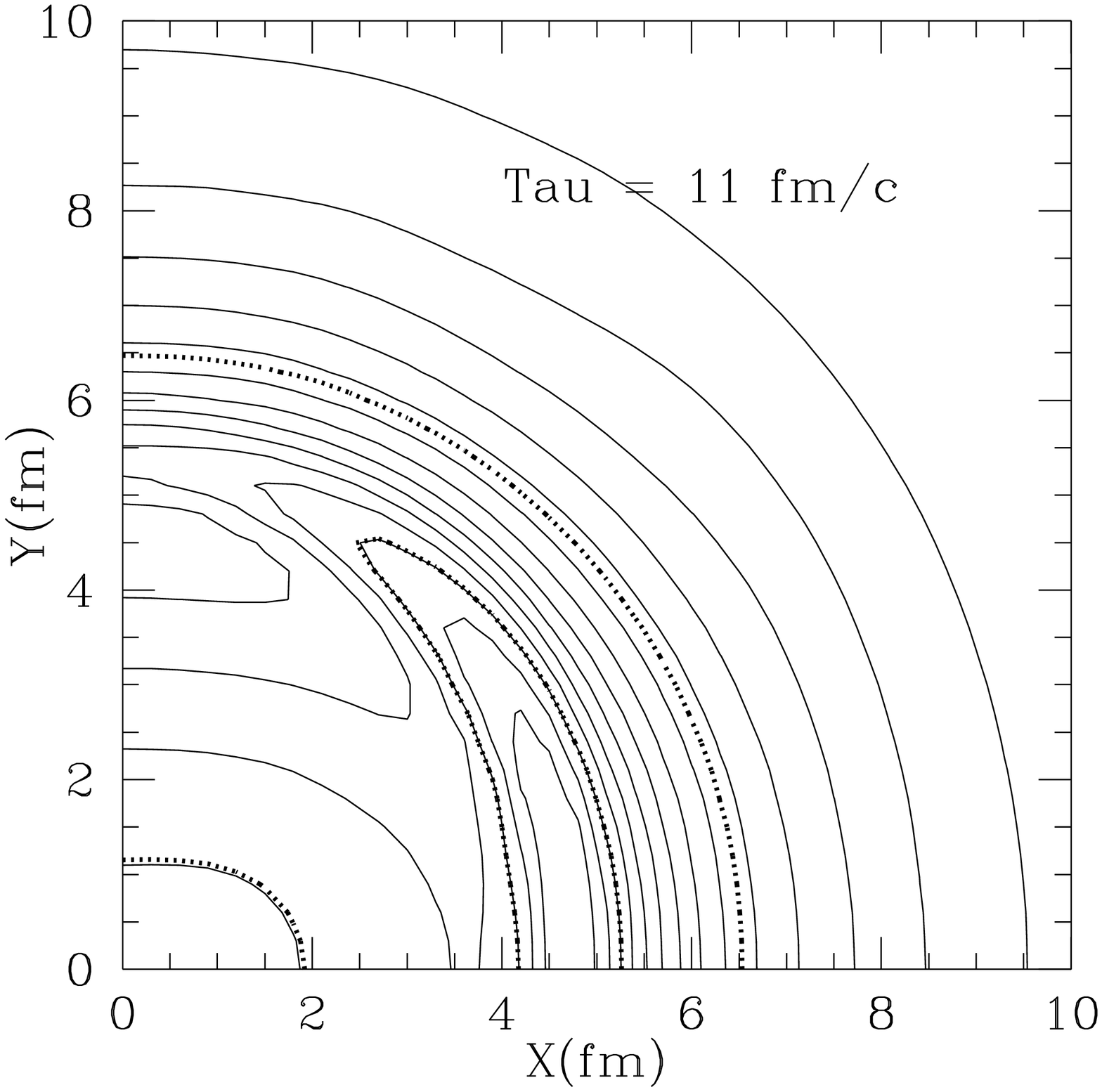,width=10cm}
\vskip -0.05in
\caption[]{
 \label{shuryak_fig2}
Typical matter distribution for AuAu collision at b=8 fm, at time t=11 fm/c,
in the transverse plane x-y, resulting from hydro calculation with lattice-inspired EOS. Transverse expansion is assumed to be rapidity-independent, while
the longitudinal expansion is Bjorken-like. The final
multiplicity assumed is $dN_{ch}/dy=850$.
Lines show levels with fixed energy density, with step ??? $Gev/fm^3$. The dotted contours are for  $T_f=120 MeV$ (the outer one), and T=140 MeV
(two inner ones).  
}
\end{figure}

\section{ RHIC as a $nutcracker$}

This section is a brief account of unusual pattern of space-time
evolution,
found for non-central collisions at RHIC energies by D.Teaney and
myself
\cite{ST}.

Let me begin with a pedagogic consideration of two opposite
schematic models of high energy  heavy ion collisions, leading to quite 
different conclusions about even such global thing as $duration$ of the
collision till freeze-out. This will set a stage for more elaborate
considerations later, based on hydrodynamic approach.

The ``model A" is just a picture of longitudinal expansion without any
transverse one, except maybe very late in the process. (One may therefore call
it a  ``late acceleration model".)
For rapidity-independent (Bjorken) expansion the dynamics
is very simple: each volume element expands linearly in proper time $\tau$.
If the total entropy S is conserved, its density $s\sim 1/\tau$. The $initial$
value of entropy density at RHIC $s_i^{RHIC}$ is of course unknown, but it
is believed to be several times that at SPS, say  $s_i^{RHIC}=(2-4) s_i^{SPS}$.
 The $final$ values should be roughly the same
(At one hand, the larger the system the more should it cool down. On the other hand, at RHIC the
fraction of baryons is expected to be significantly lower, and it should
reduce re-scattering). Therefore, this model predicts
 total duration of expansion 
$ \tau^{RHIC}=(2-4)\tau^{SPS}\sim 40 fm/c$.

The ``model B" includes the transverse expansion, but in the opposite
manner: the  observed radial flow velocity at freeze-out $v_f\approx .4$
(the value we expect to see at RHIC) is now assumed to be there all the time.
By contrast to model A, it assumes an ``early acceleration".
Including simple geometric expansion in the decrease of the entropy density
$ s \sim 1/( \tau (r_0+v_f\tau)^2)$
one finds then much shorter duration of the collision
$\tau^{RHIC}\sim 10 fm/c $ predicted by model B.

Which model is closer to reality depends on the real acceleration history,
which is in turn determined by the interplay of the 
collision geometry, the energy and the EOS. 
Qualitatively speaking, the main message of the previous section is that
while at AGS/SPS energy domain the collective flow appears late,like
in model A, at RHIC/LHC it is expected to be generated early, as in the model
B. The reason for that is specific behavior of the QCD EOS, which is 
soft in the ``mixed phase" region of energy density, $very$ soft in QGP
at $T\approx T_c$, and then rapidly becoming hard at $T\approx 2-3 T_c$.

One consequence\cite{HS}  is that duration of the 
collision $grows$ with energy in the AGS domain, but expected to $decrease$
 from SPS to RHIC. The maximal is expected to be when the initial conditions
hit the ``softest point" of the EOS, roughly at beam energy 30-40 GeV*A.
Somewhat counter-intuitive, around the same energy one expects also a maximum
value of the radial flow: longer acceleration time seem to win over softness!
Especially interesting is the dynamics of the
 ``elliptic" flow in the SPS-low RHIC 
energy domain. It is quite possible that its  energy
dependence would be sufficient to see the onset of QGP plasma.

Of course, the magnitude of the flow depends not only on hydro-EOS but also
on kinetics of the freeze-out itself. We have already mentioned two factors
which enter into consideration here: the absolute size of the system (hydro itself is scale-invariant!) and the baryon/meson ratio. Only careful
systematic study of various systems at various energies will clarify the
actual role of all these effects.

The magnitude of collective flow and its acceleration history 
 can be understood as follows. We expect
to have rapid change of pressure to energy density
ration in QGP around the phase transition. Higher density QGP has $p/\epsilon \approx 1/3$, but at the transition 
there is the minimum of this ratio (the so called ``softest point"
\cite{HS}) where  $p/\epsilon$ is small (0.1-0.05). So at AGS/SPS energies 
the expansion is slow and QGP just ``burns inward". At some point, the
outward expansion of the QGP and the inward burning may cancel each other,
leading to near-stationary ``burning log" picture \cite{RG}.
 At higher collision energies, the burning
discontinuity is  blown out  and the situation  returns to much  simpler hydro picture typical for simple EOS $p=\epsilon/3$.

 We have recently found that   the lattice-inspired EOS leads to very
unusual picture of the expansion, with quite characteristic inhomogeneous
matter distribution (to be referred below as a $nutshells$).
Stiff QGP at the center pushes against soft matter in the transition region: 
as a result some piling of matter occurs, in a shell-like structure. 
Furthermore, for non-central collisions the geometry drives expansion
more to the direction of impact parameter (called x axis) rather than y,
starting rather early. As a result, the two half-shells {\it separate}
by freeze-out, and so (at least) two separate fireballs are
actually produced. (We called this scenario a $nutcracker$.) 
Nothing like this happens for simple EOS, which always lead to
matter distribution with a maximum at the center.

  There is not much place here to display this interesting phenomenon.
In Fig.\ref{shuryak_fig2} we show a typical mater distribution.
The time 11 fm/c is around (or slightly before) the freeze-out for most matter
(it is not changed much anyway, the longitudinal expansion simply dilute it 
more). One can clearly see two shells in x direction and holes in y ones.   

How  can  such phenomenon be seen experimentally?\\
(i) We have calculated several 
 harmonics (in angle $\phi$) of flow, $v_n$.  We found quite observable
deviations from a directed+elliptic (n=1 and 2 only)
 distributions seen before up to n=6.\\
(ii) The distribution of pions can be sufficiently accurate to see it,
but with nucleons and, even better, heavier particles like deuteron-s we
find much stronger signals for ``nutcracker" scenario
in production/flow patterns. \\
(iii) Another
 dramatic changes are found if one calculates correlators used
by two-particle interferometry (HBT). 
 Strong flow plus inhomogeneous distribution
make $visible$ HBT radii to be significantly $ smaller$
to  what one might naively expect: we see only smaller
``patches" of the picture in any given direction. (By the way,
it significantly reduces conditions for momentum resolution of the detectors.).
 But for the 
same reason taking these patches all together, into a unified picture,   
is becoming more complicated.

Finally, let me emphasize it once again:
the expected ``nutcracker"
 pattern is supposed to be seen in typical non-central
events. Because most of the RHIC detectors are able to detect the impact parameter plane in most events,  there is no doubt that absolutely $any$ phenomenon,
from particle single-body distribution to $J/\psi,\Upsilon$ suppression or
``jet-quenching" would be found strongly $\phi$ dependent, if it takes place.
We will see exciting results on that, right from the first day
of RHIC operation.

\end{document}